\documentclass[longbibliography,aps,prl,superscriptaddress,reprint]{revtex4-1}

\usepackage{graphicx}
\usepackage{physics}
\usepackage{natbib}

\begin{document}


\title{Direct digital synthesis of microwave waveforms for quantum computing} 



\author{J. Raftery}
\email{james.raftery@ibm.com}
\affiliation{Department of Electrical Engineering, Princeton University, Princeton, New Jersey 08544, USA}


\author{A. Vrajitoarea}
\affiliation{Department of Electrical Engineering, Princeton University, Princeton, New Jersey 08544, USA}
\author{G. Zhang}
\affiliation{Department of Electrical Engineering, Princeton University, Princeton, New Jersey 08544, USA}
\author{Z. Leng}
\affiliation{Department of Physics, Princeton University, Princeton, New Jersey 08544, USA}
\author{S. J. Srinivasan}
\affiliation{Department of Electrical Engineering, Princeton University, Princeton, New Jersey 08544, USA}
\author{A. A. Houck}
\affiliation{Department of Electrical Engineering, Princeton University, Princeton, New Jersey 08544, USA}

\date{\today}

\begin{abstract}
Current state of the art quantum computing experiments in the microwave regime use control pulses generated by modulating microwave tones with baseband signals generated by an arbitrary waveform generator (AWG).
Recent advances in digital analog conversion technology have made it possible to directly synthesize arbitrary microwave pulses with sampling rates up to 92 gigasamples per second (GS/s).
These new high bandwidth AWG's could dramatically simplify the classical control chain for quantum computing experiments, enabling more advanced pulse shaping and reducing the number of components that need to be carefully calibrated.
Here we use a high speed AWG to study the viability of such a simplified scheme.  
We characterize the AWG and perform randomized benchmarking of a superconducting qubit, achieving average single qubit gate error rates below $5\times10^{-4}$.
\end{abstract}


\maketitle 

A multitude of promising architectures for processing quantum information have emerged, including ion traps \cite{Brown2011, Harty2014, Choi2014, Schindler2013}, spin qubits \cite{Awschalom2013, Muhonen2015}, and superconducting circuits \cite{Devoret2013, Blais2004, Wallraff2004}.
In particular, rapid progress with superconducting circuits has led to long coherence times \cite{Paik2011, Rigetti2012, Reagor2015}, fidelities approaching the fault tolerance threshold \cite{Barends2014a}, and increasingly intricate networks of qubits \cite{Kelly2015a, Riste2015b, Corcoles2015a}.
As these systems increase in complexity, the use of scalable classical control schemes will become vitally important.

\begin{figure}
\includegraphics[scale=0.5]{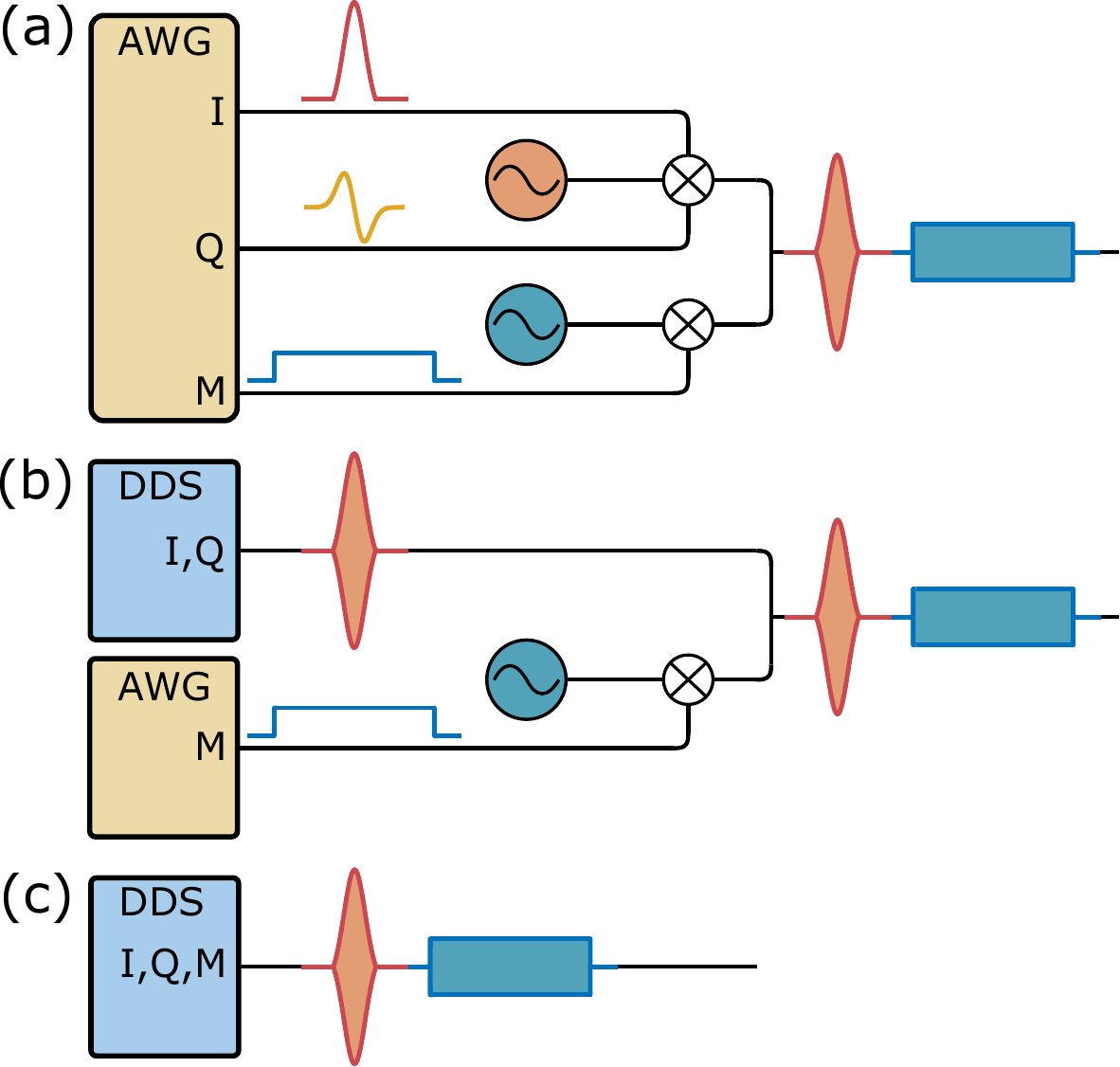}
\caption{\label{fig:upconversion} 
Three different configurations for generation of microwave control pulses for a typical circuit QED experiment.
(a) The standard procedure for upconverting control pulses into the microwave frequency range. 
The AWG generates baseband signals for the qubit pulse (I, Q, the in-phase and quadrature pulses) and a cavity readout pulse M, which are upconverted using microwave generators and mixers.
(b) In a hybrid approach, the qubit control pulses are generated via direct digital synthesis (DDS) while the measurement pulse is upconverted in the standard way. This is useful for studying the impact of DDS settings on qubit gate fidelities without effecting readout (see text).
(c) In a full DDS approach the ultra-wide bandwidth allows for generation of both the qubit and measurement pulses using the same single channel. 
}
\end{figure}

In circuit QED, characteristic qubit and cavity frequencies lie in the 4-10 GHz range.
Control pulses are typically generated at baseband by an AWG and upconverted to these microwave frequencies by mixing with a carrier.
While this standard technique works very well, upconversion does have some drawbacks.
The input chain for addressing a single qubit in an upconverting scheme invariably requires at minimum two AWG channels for quadrature modulation, a microwave generator, and a mixer.
Often in circuit QED readout pulses at the cavity frequency are sent down the same input line, requiring their own generator, mixer, and an additional power combiner.
Figure~\ref{fig:upconversion}a shows such a typical arrangement.
Care must be taken to understand the nonidealities of each component in order to attain high gate fidelities.
Mixers in particular require careful calibration - the effects of carrier leakage and other distortion products must be mitigated either through gating pulses or using single sideband modulation schemes to shift these unwanted signals away from relevant transitions \cite{Sheldon2016a}.

Advances in digital analog conversion could greatly simplify the input chain.
New high speed AWGs can operate with sampling rates as high 92 GS/s, making it possible to directly draw the microwave frequency control pulses rather than upconvert.
This direct digital synthesis (DDS) scheme has several potential benefits, which will become more pronounced as systems continue to scale up in complexity.
Most obviously, directly drawing the final microwave pulses obviates the need for separate microwave generators, presenting significant potential hardware savings.
This also allows for the elimination of mixers, dramatically simplifying the input chain and reducing the number of potential sources of infidelity.
Moreover, full dual-quadrature control can be attained via a single AWG channel.
Using the extremely wide bandwidth of these AWGs it is even possible to use a single channel to generate control pulses at multiple frequencies, addressing both qubit and cavity or even multiple qubits on a single cavity bus.
This technique is particularly attractive for situations requiring precise phase relations between signals at disparate frequencies and across multiple channels, such as multi-qubit gates \cite{Sheldon2016a}, reservoir engineering \cite{Murch2012a, Kimchi-Schwartz2016}, or the generation of interesting many-body quantum states in microwave lattices \cite{Houck2012, Hafezi2015}

\begin{figure}
\centerline{\includegraphics[scale=.5]{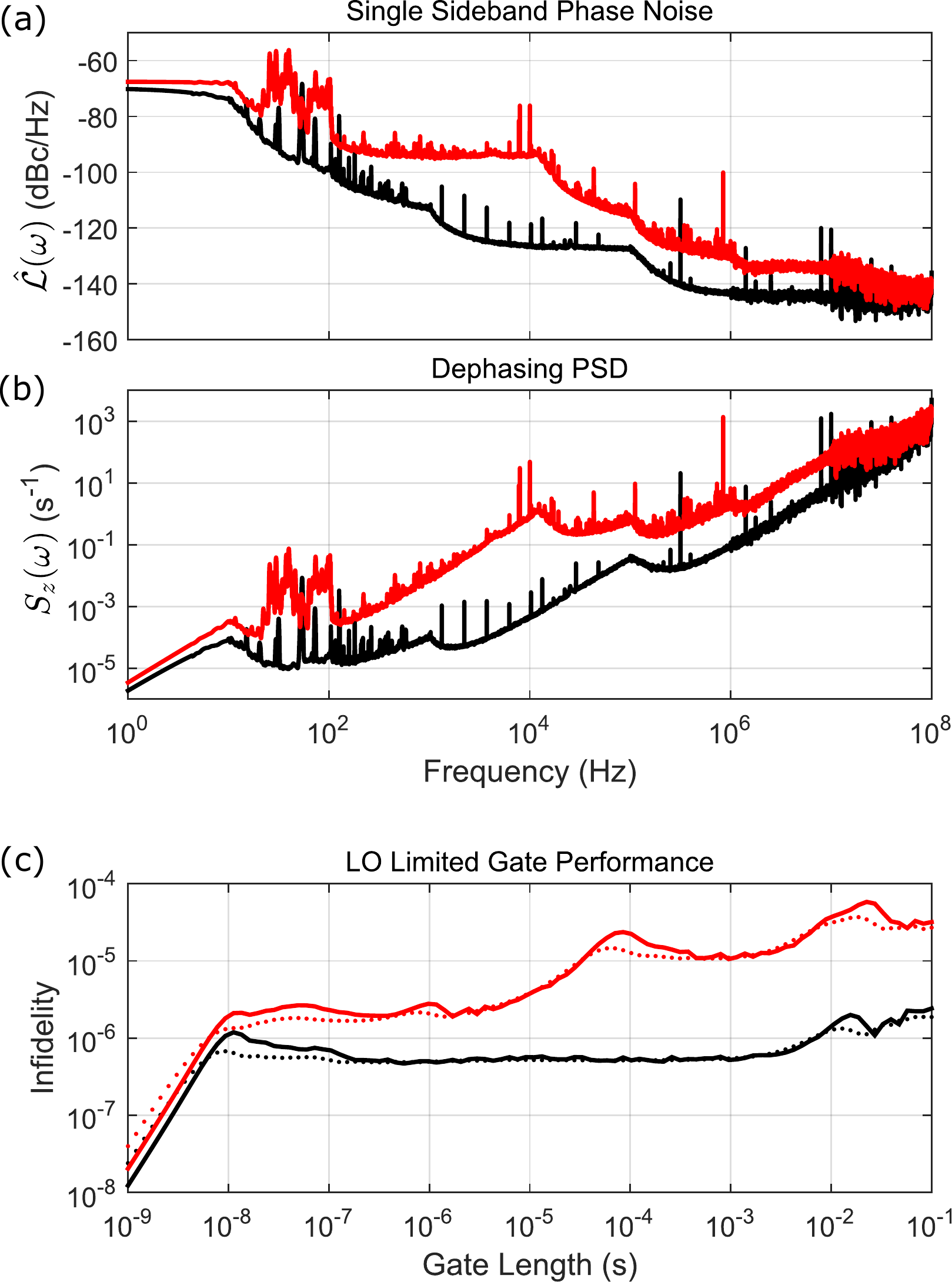}}
\caption{\label{fig:phase} 
(a) Comparison of the single sideband phase noise for a 4.77 GHz DDS microwave tone generated by a Keysight M8195a AWG (red) with the same tone generated by a Keysight E8267D PSG Vector Signal Generator (black).
A resolution bandwidth of 1 Hz was used and a linear extrapolation was used for frequencies less than 4 Hz.
(b) Unilateral dephasing power spectral density for the corresponding phase noise measurements in (a).
(c) Expected infidelities due to the dephasing PSD's in (b) as a function of control gate length. Solid lines are for an $X_{\pi}$ gate and dashed lines correspond to the identity gate.
In circuit QED single qubit gate lengths of 10-30 ns are typical, where differences in error rates between the two sources are small.
}
\end{figure}

Questions remain, however, about the viability of such an approach for quantum computing applications.
Foremost, is the quality of the generated microwave signals comparable to upconverting schemes?
It is known, for example, that phase fluctuations of the experiment's `master clock' are indistinguishable from dephasing of the qubit \cite{Ball2016}.
Since there is no separate microwave generator in a DDS setup to act as the local oscillator, the phase noise of the AWG is the directly relevant parameter.
Figure \ref{fig:phase} compares the single sideband phase noise $\hat{\mathcal{L}}(\omega)$ for a directly synthesized microwave tone at 4.773 GHz with the output of a microwave source commonly used in upconverting setups, showing significantly higher noise for the DDS tone.
Following the procedure outlined in Ball et al. \cite{Ball2016}, we convert the phase noise data to an equivalent unilateral dephasing power spectral density  (PSD)
\begin{equation}
S_z^{(1)}(\omega)=\frac{1}{2}\omega^210^{\frac{\hat{\mathcal{L}}(\omega)}{10}}
\end{equation}
Importantly, the factor of $\omega^2$ introduced when converting to the dephasing PSD leads to a decreased role of the close-in noise - where differences between the DDS system and signal generator are greatest.
We then use the dephasing PSD to estimate an infidelity floor due to master clock instability for driven and undriven cases using a filter transfer function formalism (see Supplemental Material \footnote{See Supplemental Material at [URL will be inserted by publisher] for a full wiring diagram and details of the phase noise calculations.} and Refs. \onlinecite{Ball2016, Green2012, Green2013} for details).
Despite the higher phase noise the bounds due to LO instability remain well below current and near-term limits due to qubit coherence and other sources of infidelity, implying that microwaves generated using DDS can be used as high quality local oscillators.
This analysis, however, only addresses the effective dephasing which occurs due to an unreliable reference frame, and does not evaluate the direct impact of DAC noise on the qubit.

To more concretely study the viability of DDS for quantum computing, we demonstrate low single-qubit gate error rates in a circuit QED setting.
First, a hybrid setup is used (see Figure \ref{fig:upconversion}b) to separate details of the qubit control pulses from cavity readout.
Qubit pulses are directly synthesized using a single AWG channel while the cavity readout pulse is upconverted in the conventional manner.
Experiments were performed with a single fixed frequency transmon qubit \cite{Koch2007} sample provided by IBM.
The qubit has a transition frequency of 4.773 GHz and anharmonicity of $\delta/2\pi=$ 375 MHz, and is coupled to a coplanar waveguide resonator with resonance frequency of 10.166 GHz.
$T_1$ and $T_{2 Echo}$ are $51\pm 10 \: \mu s$ and  $32\pm 1.2 \: \mu s$, as measured over 500 measurements in an 8 hour span.

A variety of techniques exist to estimate the fidelity of qubit gates, including randomized benchmarking \cite{Emerson2005a, Knill2008a, Magesan2011, Barends2014a}, quantum process tomography \cite{Chow2009, Chuang1997, Poyatos1997}, and gate set tomography \cite{Blume-Kohout2013}.
Randomized benchmarking (RB) was chosen for this experiment due to its insensitivity to state preparation and measurement (SPAM) errors and straightforward analysis.

\begin{figure}
\centerline{\includegraphics[scale=0.6]{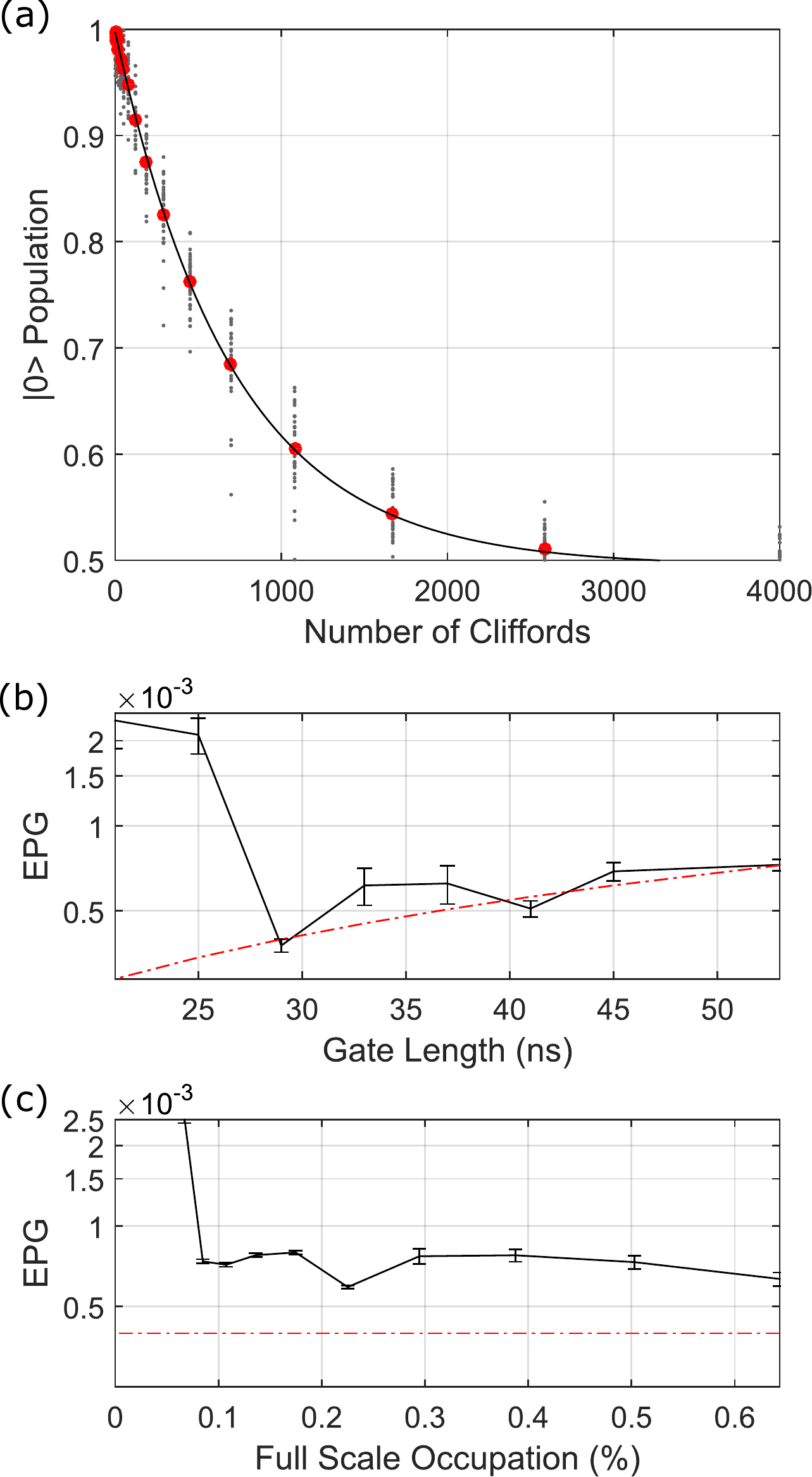}}
\caption{\label{fig:exampleRB} 
Randomized benchmarking results for directly synthesized qubit control pulses.
Clifford gates were composed of one to three $4\sigma$ gaussian pulses with DRAG correction in quadrature.
A $5\text{ns}$ buffer between primitive gates is included in the gate length.
(a) RB results for $29\,\text{ns}$ long primitive gates, where the red dots are the average of twenty random sequences (the dark gray dots).
EPG is $3.8(4)\times10^{-4}$.
(b) EPG versus gate length. 
Black data points include $\pm$ 1 standard deviation errorbars, and the red dashed line indicates the coherence limit.
(c) EPG versus fraction of the AWG full scale occupied by a tuned up $X_{\pi}$ pulse.
Gate length was fixed at $29\,\text{ns}$.
}
\end{figure}

In a single qubit RB experiment, a random sequence of $m$ Clifford operations are performed on the qubit followed by a final Clifford gate chosen to return the qubit to the ground state absent any gate errors.
Here the Clifford gates themselves are constructed from a basis set of one to three primitive gates $\left\{ I,    X_{\pm \pi/2}, X_{\pi} ,     Y_{\pm \pi/2}, Y_{\pi} \right\}$.
Following this sequence, the survival probability $P(\ket{0})$ is measured and averaged over $n$ such random sequences.
Due to the depolarizing effect of the Clifford gates, this averaged survival probability exhibits an exponential decay as a function of sequence length $m$. 
A fit to $A p^m + B$ yields an error per Clifford (EPC) of $\frac{1}{2}(1-p)$, with the $A$ and $B$ fit parameters accounting for SPAM errors \cite{Magesan2011}.
The error per primitive gate (EPG) is given by $\frac{1}{2}(1-p^{1/N_g})$ where $N_g$ is the average number of primitive gates per Clifford, 1.875 for our chosen Clifford decomposition.
A representative RB experiment is shown in Figure \ref{fig:exampleRB}a, demonstrating an EPG of $3.8(4)\times10^{-4}$.
The qubit pulses consisted of truncated gaussian pulses with DRAG-correction in quadrature \cite{Motzoi2009b}.

An important issue for direct synthesis is ensuring that the carrier phase remains continuous throughout the experiment.
In an upconverting scheme, the baseband signals can be built by concatenating many short waveform segments since the mixing process guarantees a consistent phase reference.
This can be especially useful when generating long pulse sequences out of a small library of gates preloaded onto an AWG.
When using direct synthesis, however, care must be taken when stringing together separate segments that are already upconverted, or else the carrier phase may jump and lead to an unwanted effective Z rotation of the qubit.
High memory depth allows for extremely long single waveforms even at the full sampling rate.
We generate each experiment (sequences of up to 4,000 Clifford gates) as a single waveform, constructing baseband signals for the I and Q quadratures and modulating by their respective carriers all in software before uploading the waveform to the AWG.

Automated gate tune-up procedures were implemented using error amplification schemes as in Sheldon et al. \cite{Sheldon2016a}.
Additionally, the output chain included a Travelling Wave Parametric Amplifier (TWPA) \cite{Macklin2015a}, allowing single shot measurement fidelity of 93\% and dramatically reducing the averaging time necessary for tune-up and data collection.
These automation and speed improvements made it feasible to vary pulse parameters and AWG settings while monitoring the effect on gate fidelity.
Figure \ref{fig:exampleRB}b shows EPG as a function of total gate length. 
For long pulses, the fidelity is limited by the coherence of the qubit, while leakage to noncomputational states and failure of the automated tune-up procedure limits performance for the shortest gates \cite{McKay2016a, Chen2016}.
We chose a gate length of 29 ns for further experiments as it was the fastest gate that could be reliably tuned up, allowing us to maintain low error rates while being sensitive to unitary errors of the gates rather than being coherence limited.

Addressing the qubit and cavity in a circuit QED system often requires dramatically different pulse powers, which presents a challenge for using the full DDS scheme (see Figure \ref{fig:upconversion}c).
In this sample the significant detuning between qubit and cavity lead to a calibrated $X_{\pi}$ pulse requiring approximately 45 dB more power than the readout pulse at the top of the fridge, despite significant filtering at the cavity frequency inside the fridge (see Supplement for a full fridge wiring diagram).
In upconverting schemes the power for each microwave generator (more specifically the amount of attenuation used after mixing) can be individually chosen to optimize each pulse.
This highlights an important tradeoff for direct synthesis schemes, where generating different types of pulses on a single channel sacrifices this flexibility.
As the AWG only has a vertical resolution of 8 bits, too large a difference would force the weaker measurement pulse to occupy a small fraction of the dynamic range and lead to significant quantization error.
Filtering on the input line was carefully chosen to ensure comparable pulse amplitudes at the output of the AWG.

To explore this issue we studied gate error rates as a function of the AWG full scale using 29 ns pulses (see Figure \ref{fig:exampleRB}c).
The system was used in a hybrid DDS setup so that the full scale settings could be changed without effecting the readout pulse.
 Error remained low all the way down to tuned up $X_{\pi}$ pulse amplitudes as low as 8.49\% of the full scale.
 For this amplitude, the corresponding $X_{\pi/2}$ pulse amplitude was only 4.47\%, and the drag component was only 0.96\%, corresponding to only 4.5 and 2.3 bits of resolution respectively.

The effect of sampling rate on gate fidelity was also studied.
The internal digital analog converters (DACs) of the AWG can be set to have a sampling rate between 57.76-65 GS/s.
Further, the hardware can be run in a variety of different modes that impact the overall sampling rate, halving or even quartering the available sampling rate depending on the number of channels being used. 
Even using the lowest sampling rate possible, 14.44 GS/s, randomized benchmarking experiments did not show any significant reduction in qubit gate fidelity.

After studying the quality of the qubit control pulses in isolation, we also examined the performance of the system when in full DDS mode (See Figure \ref{fig:upconversion}c).
The cavity pulse was generated 120 ns after the final qubit gate using the same single channel of the AWG.
Gates were tuned up as normal, and no significant differences were noted for short sequences of qubit pulses or for readout fidelity.
For long sequences, however, we observed a small but measurable reduction in the amplitude of the cavity pulse.
The size of this effect increased monotonically with the length of the qubit pulse sequence.
For a randomized benchmarking experiment this manifested as a sequence length dependent distortion and lead to the decay of the fidelity signal to a value of .6 rather than the expected .5.

Further investigation revealed a hardware limitation completely independent of the circuit QED experiment, and the effect could be observed just measuring the pulses generated from the DDS AWG directly.
The presence of a long pulse at one frequency distorts an immediately following pulse.
Varying the delay between the two pulses showed a $31 \mu s$ exponential decay time for the distortion.
Varying the length of the first pulse shows the distortion saturates with a similar timescale, such that the effect is noticeable only for long series of pulses such as randomized benchmarking, or experiments involving long sustained drives like a Rabi type experiment.

One workaround is to account for the distortion in a background measurement that is subtracted off.
For every sequence, a corresponding background measurement is taken where the exact same qubit control pulses are applied, but shifted a few GHz from the qubit frequency.
While the sequence applied in the background measurement does not excite the qubit or effect the cavity, it does create the same sequence dependent distortion.
Randomized benchmarking experiments utilizing full DDS with this background subtraction routine show the proper decay to .5 and the same high fidelities observed using the hybrid DDS/upconversion method.

In this work we have studied the viability of DDS as an alternative technique for generation of microwave controls for circuit QED experiments.
This technique is significantly simpler than conventional upconversion techniques - requiring no mixing and fewer hardware channels.
While phase noise for the measured DDS AWG is higher than the rf-generators used in our current upconverting setups, we do not expect this to limit fidelity due to master clock dephasing for near-term experiments in circuit QED.
Randomized benchmarking experiments demonstrate that DDS systems can indeed generate high fidelity qubit control pulses with average gate error rates less than $5\times10^{-4}$. 
Importantly, a hardware limitation was identified when using the system in a full DDS mode, leading to a sequence length dependent distortion of the measurement pulse.
While the distortion did not affect randomized benchmarking fidelities, it is important for further studies to be carried out to identify whether this hardware limitation is fundamental to DDS systems or if it is peculiar to the equipment tested here.



%
%

%

\begin{acknowledgments}
The authors would like to thank IBM for sample fabrication and MIT Lincoln Labs for providing the TWPA for this experiment.
This work is supported by the Department of Defense under contract W911NF-15-1-0421 and IARPA under contract W911NF-10-1-0324.
\end{acknowledgments}


%

\end{document}